\documentclass[aps,prb,groupedaddress,twocolumn,showpacs]{revtex4}
\usepackage{psfig,graphics,graphicx}
\usepackage{amssymb}
\usepackage{bookmath}
\begin{document}
\title{Crystalline order in superfluid $^3$He films}
\author{A. B. Vorontsov}
\author{J. A. Sauls}
\affiliation{Department of Physics and Astronomy,
             Northwestern University, Evanston, IL 60208}
\date{\today}

\begin{abstract}
We predict an inhomogeneous phase of superfluid \He\
films in which translational symmetry is spontaneously broken in the
plane of the film.
This phase is energetically favored over a
range of film thicknesses, $D_{c_2}(T)<D<D_{c_1}(T)$, separating
distinct homogeneous superfluid phases. The instability at the
critical film thickness, $D_{c_2}\approx 9\,\xi(T)$, is a
single-mode instability generating striped phase order
in the film. Numerical calculations of the order parameter and free
energy indicate a second-order instability to a periodic lattice of
degenerate B-like phases separated by domain walls at
$D_{c_1}\approx 12\,\xi(T)$. The striped phase should be identifiable
in transport and nuclear magnetic resonance experiments.
\end{abstract}

\maketitle
The phases of superfluid \He\ provide a beautiful example of
spontaneously broken symmetry in condensed matter physics,
exhibiting properties common to superconductors, nematic liquid
crystals and anti-ferromagnets.
Many of the unique physical properties of superfluid \He,
including the spectrum of low-energy excitations, are connected
to the spontaneous breaking of orbital and spin rotation
symmetries in combination with global gauge symmetry that is
associated with superfluidity and superconductivity.
Inspite of the complex order that develops, the bulk \textsf{A}
and \textsf{B} phases of $^3$He are translationally invariant.
Indeed translational symmetry is generally
assumed to hold even in reduced dimensions, e.g. superfluid
films.\cite{li88,nag00}

NMR measurements on relatively thick
($\mu\text{m}$) films show evidence of an \textsf{A}- to \textsf{B}-like
transition predicted within the context of Ginzburg-Landau (GL)
theory.\cite{kaw98}
However, unexplained anomalies in film flow\cite{xu90} and third sound
experiments\cite{sch98} suggest that our current theoretical
understanding of the phases of superfluid $^3$He films is
insufficient. One of the intriguing questions raised by these
experiments is whether or not there may be qualitatively new phases
stabilized in reduced dimensions.\cite{vor04}

Here we report the theoretical prediction of a phase of superfluid
\He\ exhibiting \textsl{spontaneously} broken translational
symmetry, i.e. crystalline order. This phase is shown 
theoretically to be the stable ground
state of a superfluid \He\ film, with the broken translational
symmetry occurring in the plane of the film. The mechanism
responsible for this phase is competition between surface depairing
and domain wall formation between degenerate ground states, and is
generic to \He\ confined in at least one spatial dimension.

\par
The superfluid phases of $^{3}$He are Bardeen-Cooper-Schrieffer
(BCS) condensates of orbital p-wave ($L=1$) Cooper pairs formed
from quasiparticles with zero total momentum ($+\vp,-\vp$) near the
Fermi surface in spin-triplet ($S=1$) states.
In terms of the basis of triplet states the order parameter is given by
{\small
\be
\vDelta
  = \Delta_{+}(\hat{\vp})\ket{\uparrow\uparrow}
   +  \Delta_{-}(\hat{\vp})\ket{\downarrow\downarrow}
  + \Delta_{0}(\hat{\vp})\frac{1}{\sqrt{2}}
                 \ket{\uparrow\downarrow+\downarrow\uparrow}
\,,
\ee
}where
$\Delta_{m}(\hat{\vp})=\sum_{i=x,y,z}\textsf{A}_{mi}\,\hat{\vp}_i$
for $m=0,\pm 1$. There are two bulk phases of superfluid \He\
in zero field. For a narrow temperature range near $T_c$
at high pressures, $p>p_c=21\,\text{bar}$, \He\ condenses into
the \textsf{A} phase with an order
parameter of the form, $\Delta_{+}=\Delta_{-}=0$ and
$\Delta_{0}=\Delta(T)\left(\hat{\vp}_x+i\hat{\vp}_y\right)$.
This phase exhibits anti-ferromagnetic spin correlations, and
an orbital state that breaks time-inversion symmetry, i.e. a
condensate of pairs with orbital angular momentum $+\hbar$.
The \textsf{B}-phase, which is the stable state over most of
the phase diagram in zero magnetic field, is a superposition
of all three triplet spin states and all three orbital
states, with
$\Delta_{+}=\Delta(T)\left(\hat{\vp}_x-i\hat{\vp}_y\right)/\sqrt{2}$,
$\Delta_{-}=\Delta(T)\left(\hat{\vp}_x+i\hat{\vp}_y\right)/\sqrt{2}$,
$\Delta_{0}=\Delta(T)\hat{\vp}_z$. This state describes a
condensate of spin-triplet, p-wave pairs in a state with
total angular momentum $J=0$. There is a continuous manifold
of B-phase states related by a relative rotation of the spin
and orbital coordinate axes. Surface and nuclear dipolar
energies resolve most, but not all,
of the degeneracy.
In addition to the bulk \textsf{A}- and \textsf{B}-phases, the planar
(\textsf{P}) phase is a possible ground state for thin films of
$^3$He. The \textsf{P}-phase is a
two-dimensional version of the \textsf{B}-phase with $\Delta_0=0$.
Alternatively, the \textsf{P}-phase is an equal amplitude superposition of
degenerate, time-reversed \textsf{A}-phase orbital states with opposite 
angular momenta. As a result the \textsf{P}-phase is degenerate with the
\textsf{A}-phase in the weak-coupling BCS theory, but preserves time-inversion
symmetry.

\par
Here we consider \He\ films of uniform thickness, $D$, bound to a solid substrate.
The liquid-vapor interface is assumed to be perfectly reflecting and atomically
smooth. Thus, we consider $p\rightarrow 0\,\text{bar}$. This is also the
weak-coupling limit for superfluid $^3$He, as indicated by
the heat capacity jump $\Delta C/C_N\rightarrow 1.43$ for $p\rightarrow
0\,\text{bar}$.\cite{gre86}
Substrates may provide a range of scattering from specular to diffuse
scattering depending on the degree of roughness. We consider specular and fully
diffuse scattering using boundary conditions described in Ref. \onlinecite{vor03}.

The order parameter for the film geometry is defined in terms of $x$- and
$y$-axes which lie in the plane of the film and the $z$-axis perpendicular
to the film. Scattering of $^3$He quasiparticles off the free surface and
substrate suppresses the orbital $\hat{\vp}_z$-component of the order
parameter (for either specular or diffuse scattering) in films less than
about $1\mu$m thick. In such thin films the \textsf{A}-phase or the
\textsf{P}-phase is stable; in the weak-coupling limit these states are
degenerate even with strong pairbreaking from diffuse scattering.
In thin films the excitation spectrum is typically dominated by gapless excitations.
The effect of this spectrum on strong-coupling energies, combined with relative
importance of gradient energies for low-temperature, low-pressure thin films means
that we cannot infer the relative stability of phases from what is known about
strong-coupling energies in bulk \He.

\begin{figure}[t]
\vspace*{5mm}
\centerline{\includegraphics[width=\hsize]{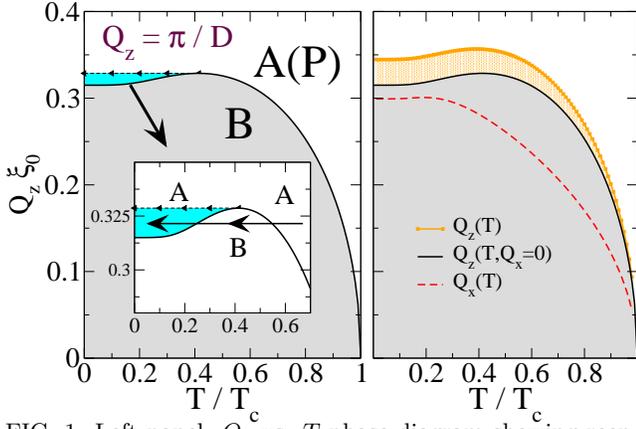}}
\vspace*{-5mm}
\caption{Left panel: $Q_z$ vs. $T$ phase diagram showing reentrance for homogeneous
         phases. Right panel: Instability onsets for higher $Q_z(T)$
         [orange line with boxes] for an inhomogeneous phase
         with in-plane wavevector $Q_x(T)$ [red dashed curve].
         Note that
         $\xi_0 = \hbar v_F / 2\pi T_c = 77$ nm at $p=0$ bar.}

\label{fig:Q_vs_T}
\end{figure}
Microscopic calculations show that as the film thickness
increases surface scattering is unable to completely suppress the
$\hat{\vp}_z$-component of the order parameter. Results by
several groups predict that equilibrium phase for film thickness,
$D\gtrsim D_c(T)\sim 10\,\xi(T)$, is the deformed \textsf{B} phase described by the order
parameter,
$\vDelta_{B}=(\Delta_\parallel\hat{\vp}_x,\Delta_\parallel\hat{\vp}_y,\Delta_z\hat{\vp}_z)$,
\footnote{Hereafter we use the more compact notation for the spin components:
$\vDelta=(\Delta_x,\Delta_y,\Delta_z)\equiv\Delta_x\ket{x}+\Delta_y\ket{y}+\Delta_z\ket{z}$,
with the cartesian spin basis defined by
$\ket{x}=(-\ket{\uparrow\uparrow}+\ket{\downarrow\downarrow})/\sqrt{2}$,
$\ket{y}=i(\ket{\uparrow\uparrow}+\ket{\downarrow\downarrow})/\sqrt{2}$,
$\ket{z}=\ket{\uparrow\downarrow+\downarrow\uparrow}$.
}
with $\Delta_z=\Delta_{\perp}(T)\sin(\pi\,z/D)$.\cite{har88,nag00,vor03}
The transition is first- or second-order at a critical film thickness,
$D_c(T)$, depending on whether the low temperature phase is the
\textsf{A}-phase or the \textsf{P}-phase. The phase boundary, taken from
our earlier calculation,\cite{vor03} is shown in the left panel of Fig.
\ref{fig:Q_vs_T} in terms of the critical wavevector, $Q_z(T)=\pi/D_c(T)$.
The inset emphasizes the re-entrance ($A\leftarrow B\leftarrow A$) for
$T\lesssim 0.42\,T_c$ near the critical line, which suggests that a lower
energy state at low temperatures, in the vicinity of the critical line,
may be achieved by an inhomogeneous phase that incorporates features of
both phases. This is the case, but as we show below the structure of
inhomogeneous phase is more complex than any of the homogeneous phases and
evolves over a relatively wide range of film thickness.

\par
The transition from B-phase to the P-phase is second-order on the critical line,
$D_c(T)$. Thus, we first look for a second-order instability to an inhomogeneous
phase that pre-empts the P-B transition. Our starting point is the weak-coupling
gap equation,
\bea
\lefteqn{
\onethird\ln\left(\frac{T}{T_c}\right)\,\vDelta^{(\pm)}(\hat{\vp},\vR)=
\int{d\Omega_{\hat{p}'}\over 4\pi}(\hat{\vp}\cdot\hat{\vp}')}&& \nonumber\\
&&\times T\sum_{m}\,
\left(\vf^{(\pm)}(\hat{\vp}',\vR;\varepsilon_m)-
\pi\frac{\vDelta^{(\pm)}(\hat{\vp}',\vR)}{|\varepsilon_m|}\right)
\,,
\label{eq:scf}
\eea
where $\vDelta^{(\pm)}(\hat\vp,\vR)$ are the real $(+)$ and imaginary $(-)$ parts
of the order parameter, and
$\vf^{(\pm)}(\hat\vp,\vR;\varepsilon_m)=[\vf(\hat\vp,\vR;\varepsilon_m)
\pm\vf(\hat\vp,\vR;-\varepsilon_m)^*]/2$
are the corresponding pair propagators in the Matusbara formulation
for equilibrium Fermi superfluids.\cite{ser83,vor03}
These objects satisfy second-order mode equations
with the order parameter providing source terms,
{\small
\begin{widetext}
\bea
\hspace*{-10mm}
\onefourth(\vv_f\cdot\grad)^2\,\vf^{(+)}-\omega_m^2\,\vf^{(+)}=
-\pi\,\omega_m\,\left[\vDelta^{(+)}+\vdelta^{(+)}\right]+\frac{\pi}{\omega_m}
&\left[\right.&
\vDelta^{(+)}(\vDelta^{(+)}\cdot\vdelta^{(+)})
+ \vDelta^{(-)}(\vDelta^{(+)}\cdot\vdelta^{(-)})
\nonumber \\
&&- \left. \vDelta^{(-)}\times(\vDelta^{(-)}\times\vdelta^{(+)})
+   \vDelta^{(+)}\times(\vDelta^{(-)}\times\vdelta^{(-)})
\right]\,,
\label{eq:f+}
\\
\hspace*{-10mm}
\onefourth(\vv_f\cdot\grad)^2\,\vf^{(-)}-\omega_m^2\,\vf^{(-)}=
-\pi\,\omega_m\,\left[\vDelta^{(-)}+\vdelta^{(-)}\right]+\frac{\pi}{\omega_m}
&\left[\right.&
\vDelta^{(-)}(\vDelta^{(-)}\cdot\vdelta^{(-)})
+  \vDelta^{(+)}(\vDelta^{(-)}\cdot\vdelta^{(+)})
\nonumber \\
&&-  \left.\vDelta^{(+)}\times(\vDelta^{(+)}\times\vdelta^{(-)})
+\vDelta^{(-)}\times(\vDelta^{(+)}\times\vdelta^{(+)})
\right]\,.
\label{eq:f-}
\eea
\end{widetext}
}
Note that $\vv_f=v_f\hat{\vp}$ is the Fermi velocity, $\varepsilon_m=(2m+1)\pi T$
is the Matsubara energy, and
$\omega_m^2\equiv\varepsilon_m^2+|\vDelta^{(+)}|^2+|\vDelta^{(-)}|^2$, where
$\vDelta^{(\pm)}$ is the order parameter of the unperturbed, translationally
invariant phase. Lastly, $\vdelta^{(\pm)}$ is the first-order correction we
seek to find.
These equations are valid up to first order in order parameter corrections,
and in their derivation we assumed that the gradient terms of $\vf^{(\pm)}$
in strongly confined space are of the same order as $\vf^{(\pm)}$ themselves.

For the P-state we can fix the overall phase so that $\vDelta$ is real. We
then have $\vDelta^{(-)}=0$ and
$\vDelta^{(+)}=\Delta_{\parallel}(z)\left(\hat{\vp}_x,\hat{\vp}_y,0\right)$.
The instability to an inhomogeneous phase is then a single-mode instability
for pairs with zero spin projection along $\vz$.
\footnote{We focus on instabilities of the P-phase since a second-order,
continuous transitions connecting to an intermediate phase that evolves
into the B-phase is possible. Our analysis for the A-phase shows no
second-order instabilities to an inhomogeneous phase.}
The eigenfunction for the instability has the form,
\be
\delta_z(\hat{\vp},\vR)=
\sum_{j=x,y,z}\,a_{z,j}(\vQ)\,e^{i\vQ\cdot\vR}\,\hat{\vp}_j
\,,
\label{eq:mode_expansion}
\ee
For a single mode instability in the plane of the film we can choose $\vQ=(Q_x,0,Q_z)$.
The resulting solution for the Fourier component of the $S_z=0$ pair propagator is,
\be
f^{(+)}_{z}(\hat{\vp},\vQ) =
\pi\omega_m(\hat{\vp})
\frac{\sum_{i}\,a_{z,i}(\vQ)\,\hat{\vp}_i}
     {\onefourth(\vv_f\cdot\vQ)^2+\omega^2_m(\hat{\vp})}\,
\,.
\nonumber
\ee

Application of the boundary conditions to the pair
propagator at the free surface ($z=D$) and substrate
($z=0$) yield a set of eigenvector equations for the
unstable mode. Here we discuss only specular scattering by
the substrate, in which case the boundary condition is,
\be
\vf^{(\pm)}(x,z=0,\hat{\vp})=\vf^{(\pm)}(x,z=0,\hat{\ul{\vp}})
\ee
for any $x$ and $\hat{\vp}$, with
$\hat{\ul{\vp}}=\hat{\vp}-2\hat{\vn}(\hat{\vn}\cdot\hat{\vp})$;
and similarly for the free surface.
The boundary conditions reduce to connections between the
Fourier components of the order parameter,
$
a_{z,x}(Q_x, -Q_z)=+a_{z,x}(Q_x, Q_z)\,,
a_{z,y}(Q_x, -Q_z)=+a_{z,y}(Q_x, Q_z)\,,
a_{z,z}(Q_x, -Q_z)=-a_{z,z}(Q_x, Q_z)\,,
$
and fixes the wavevector $Q_z = \pi/D$ in terms of the film thickness
$D$ at the instability. These results and the gap equation generate the
eigenvalue equations for the mode amplitudes, $a_{z,i}(\vQ)$,
\be
\ln(T/T_c)\,a_{z,i} - \sum_{j=x,y,z}\,I_{ij}\,a_{z,j} = 0\,\,\,,\,\,\,i=x,y,z
\,,\label{eq:eigenvalue_equation}
\ee
with the matrix elements given by
{\small
\be
\hspace*{-2mm}
I_{ij} = 6\pi T\sum_{m=0}^{\infty}\int\dangle{p}\,\hat{\vp}_i\,\hat{\vp}_j\,
\left(
\frac{\omega_m}
{\onefourth (\vv_f\cdot\vQ)^2+\omega^2_m}
-\frac{1}{\varepsilon_m}
\right)
\,.
\ee
}
\hspace*{-3mm}
Translational symmetry is unbroken along the $y$-axis in which case
$I_{xy}=I_{yz}=0$. The mode amplitudes separate into linearly
independent blocks: a 2D $(a_{z,x},a_{z,z})$ and a 1D $(a_{z,y})$
block. A non-trivial solution to Eq.~\ref{eq:eigenvalue_equation}
exists if, $(\ln(T/T_c)-I_{yy})=0$ or
$\{(\ln(T/T_c)-I_{xx})(\ln(T/T_c)-I_{zz})-I_{xz}^2\}=0$. The
eigenvalue equation for the 1D mode amplitude has a maximum unstable
wavevector only for the transition to the homogeneous phase,
$Q_z=\pi/D_c(T)$, $Q_x=0$. However, the eigenvalue equation for the 2D
block gives an unstable mode $Q_z(Q_x,T)$ that pre-empts the
homogeneous transition. The maximum value of $Q_z(Q_x,T)$ as a
function of $Q_x$ for each temperature determines the \textsl{lower}
critical film thickness, $D_{c2}(T)<D_c(T)$, for the transition to an
inhomogeneous film with broken translational symmetry in the plane of
the film. The critical wavevector, $Q_z(T)$, and the locus of values
of $Q_x(T)$ are shown in the right panel of Fig.(\ref{fig:Q_vs_T}).

\begin{figure}[h]
\centerline{\hspace*{5mm}\includegraphics[width=0.45\hsize]{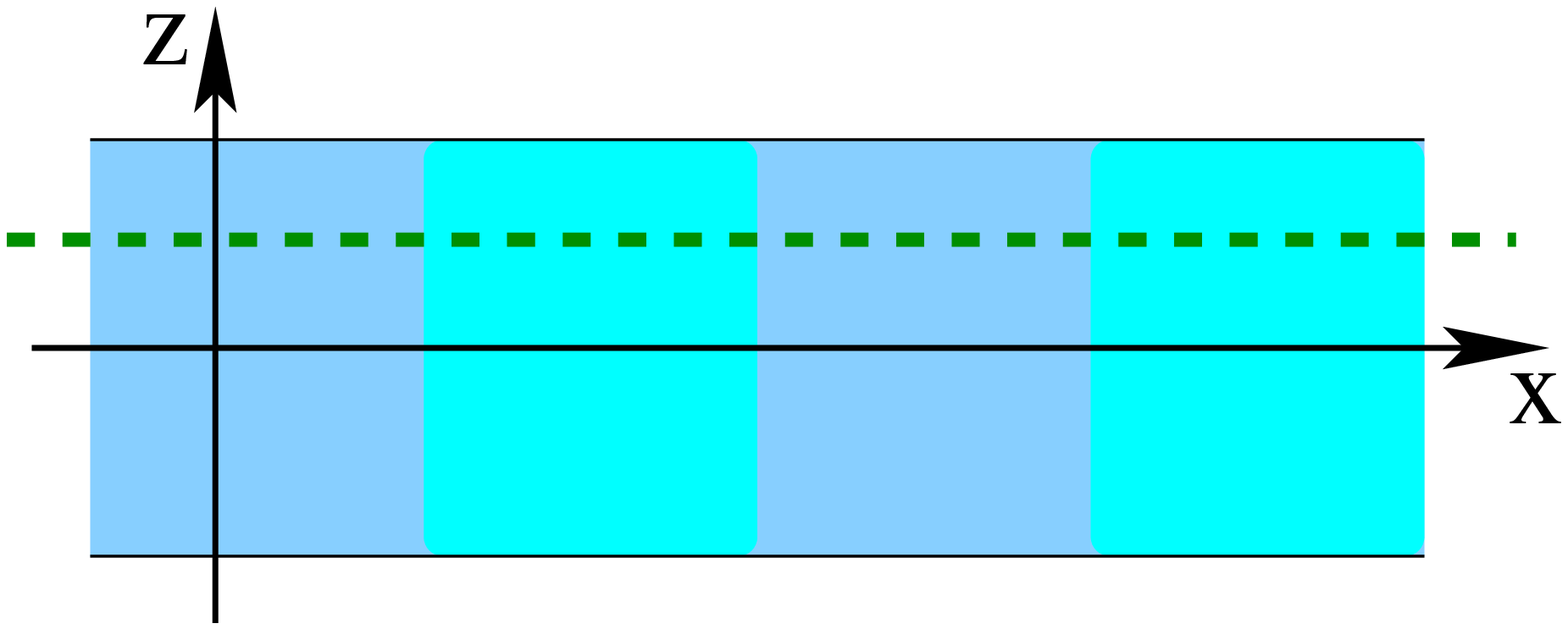}
            \includegraphics[width=0.45\hsize]{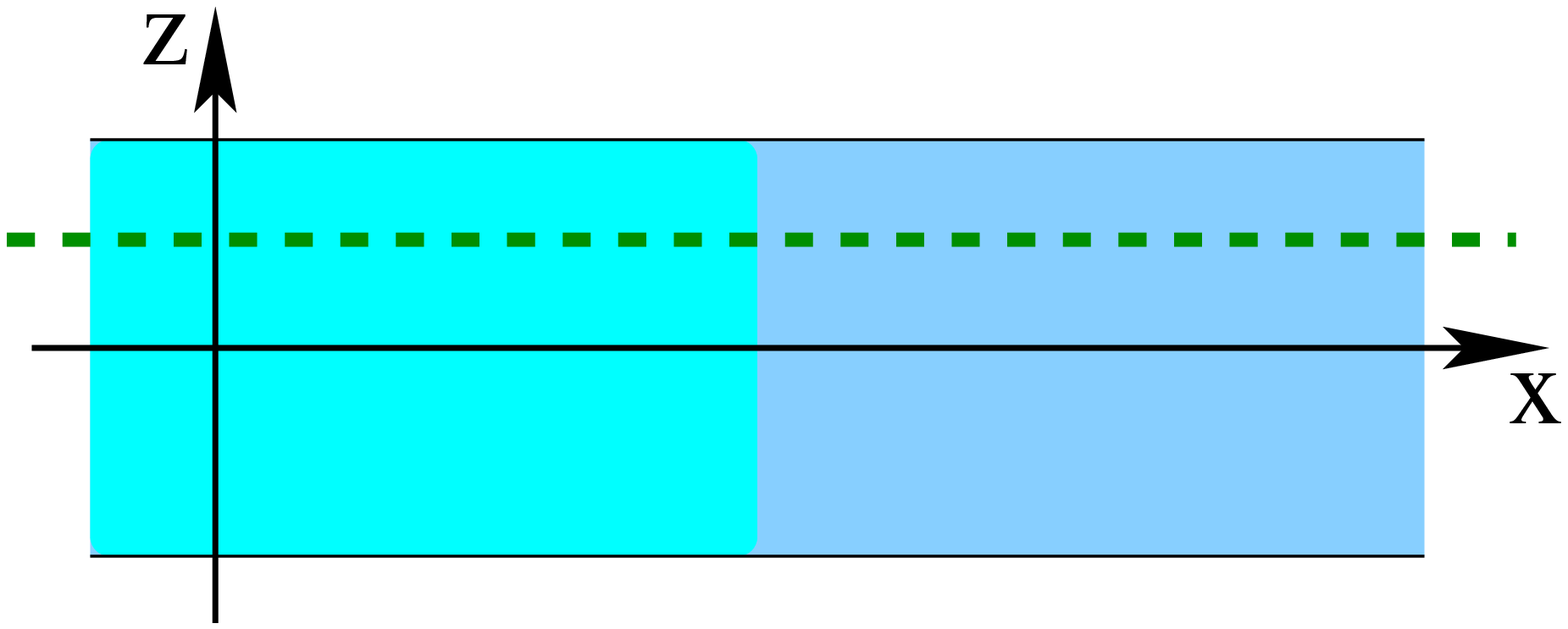}}
\centerline{\includegraphics[width=0.45\hsize]{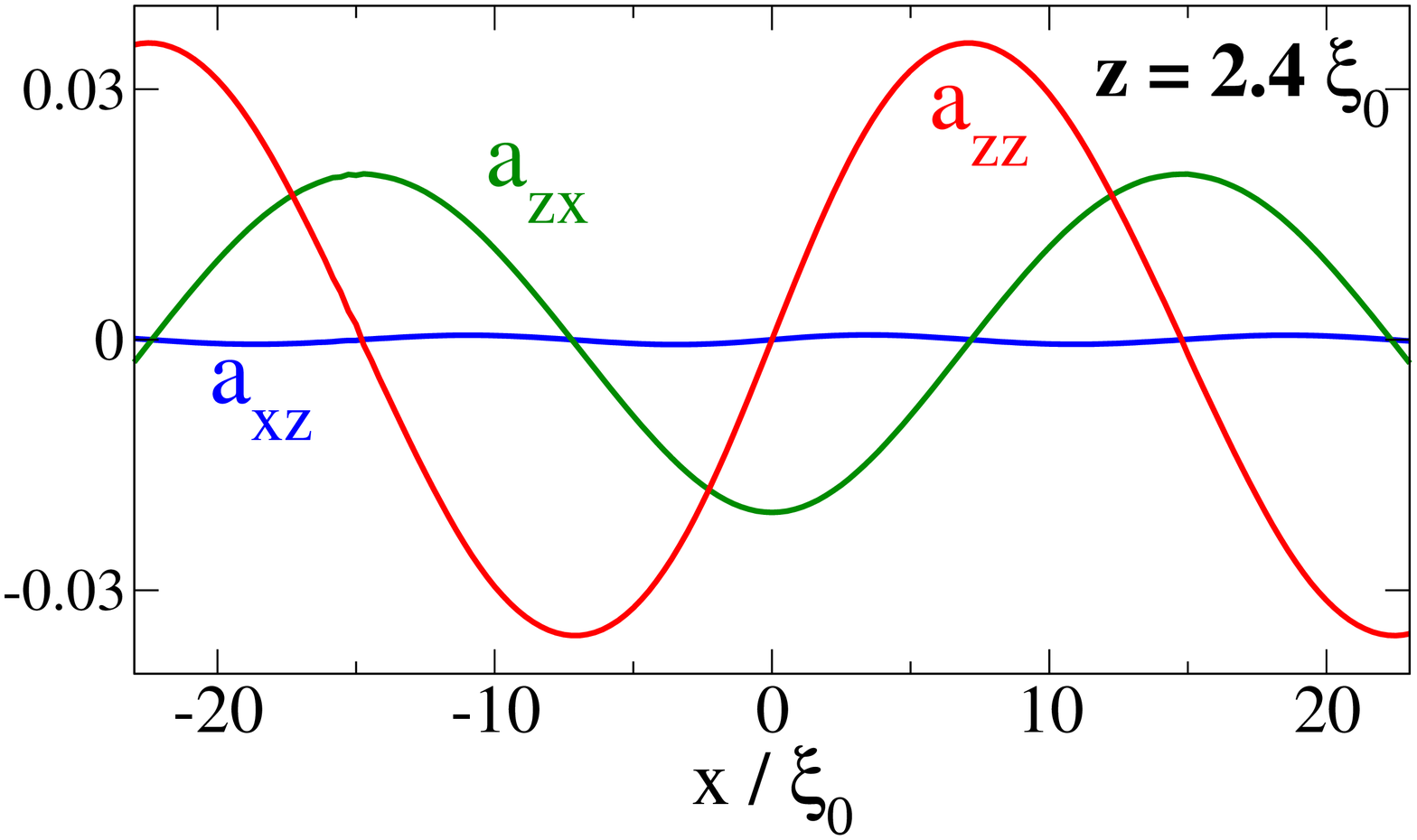}
            \includegraphics[width=0.45\hsize]{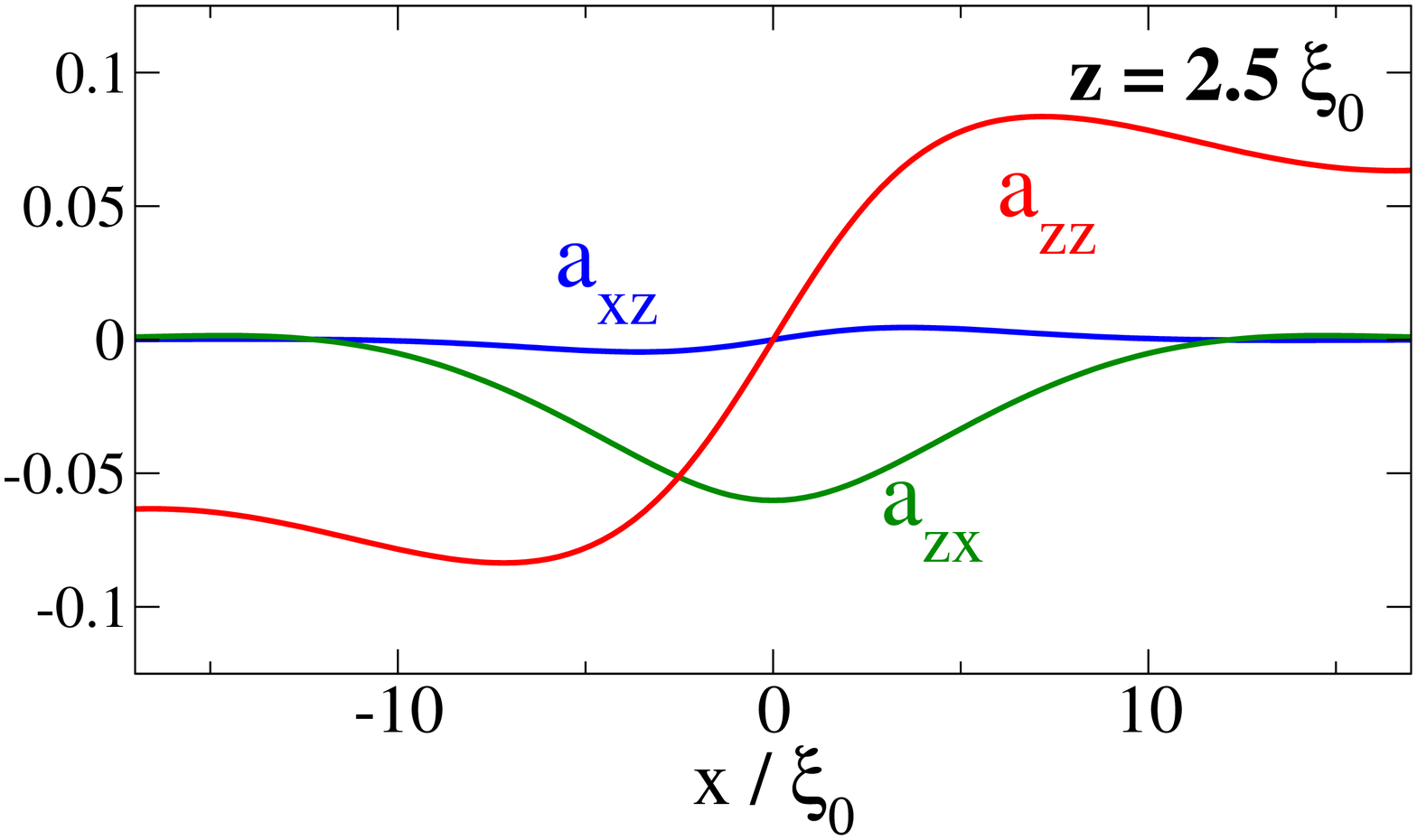}}
\caption{Order parameter amplitudes (in units of $2\pi T_c$)
         for the stripe phase at $T=0.5T_c$ along the film
         for $z\approx 2.5\xi_0$.
         Left panel: $D_{c2} < D=9.3\,\xi_0 \lesssim D_c$.
         Right panel: $D_c < D=10\,\xi_0 <D_{c1}$.
}
\label{fig:mode_amplitudes}
\end{figure}

The key signature of spontaneously broken translational symmetry
the $xy$-plane is the appearance of the order parameter amplitudes,
$a_{z,x}(Q_x,Q_z)\exp(iQ_x x)\cos(Q_z z)$ and
$a_{z,z}(Q_x,Q_z)\exp(iQ_x x)\sin(Q_z z)$. These amplitudes are shown in
the left panel of Fig. \ref{fig:mode_amplitudes} for $T=0.5\,T_c$ and
$D=9.3\,\xi_0 \lesssim D_{c}(T)$.
Note that the full solution for the order parameter
above the lower critical thickness also shows very small oscillatory amplitudes for
the in-plane spin-components, e.g. $a_{x,z}$.

Modes with the same magnitude for the unstable wavevector, $Q_{x}$,
but different orientation in the plane of the film are degenerate.
In the absence of an external bias to select the direction of
the unstable mode, the instability may propagate in any direction
in the plane of the film. For $D>D_{c2}$ the spatial structure of the
order parameter that is realized is determined from the minimum free energy.
This phase may exhibit one-dimensional, stripe-phase order, or possibly
a two-dimensional structure defined by two non-collinear wavevectors, e.g.
a triangular lattice. A comparison of the possible minimum energy
configurations of the inhomogeneous phase has not been carried out. Here
we focus on the structure of the one-dimensional stripe phase.

The broken symmetry phase persists for film thickness
well above the original critical line, $D_c(T)$, for the homogeneous A-B
transition. For thicker films, actually for $D>D_{c2}(T)$, the gap equation
includes nonlinear driving terms that couple modes with different wavevectors. The
ground state is periodic, but the structure is non-sinusoidal. The right panel of
Fig. \ref{fig:mode_amplitudes} shows the order parameter amplitudes for a film
with $D=10\,\xi_0 > D_{c}(T)$. The basic structure of this phase is indicated by
the amplitude $a_{z,z}$, which has developed a soliton-like structure separating
``domains'' of degenerate B-like phases: e.g.
{\small $\vDelta_{B}^{<}=(\Delta_{\parallel}\hat{\vp}_x,\Delta_{\parallel}\hat{\vp}_y,-\Delta_{\perp}\hat{\vp}_z)$}
and
{\small $\vDelta_{B}^{>}=(\Delta_{\parallel}\hat{\vp}_x,\Delta_{\parallel}\hat{\vp}_y,+\Delta_{\perp}\hat{\vp}_z)$}.
Also, centered on the soliton is a non-B-like phase, represented by
$a_{z,x}$, bound to the domain wall.

\begin{figure}[h]
\centerline{\includegraphics[width=0.85\hsize]{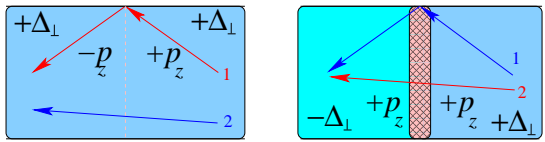}}
\caption{Left panel: Surface reflection (trajectory 1) with
$p_z\rightarrow-p_z$ leads to strong pairbreaking for
the homogenous B-phase. Right panel: Trajectory 2 is a the strong pairbreaking
trajectory because of the sign change $\Delta_{\perp}\rightarrow -\Delta_{\perp}$.
}
\label{fig:scattering}
\end{figure}

This basic structure also provides a clue to the underlying mechanism
stabilizing the inhomogeneous phase; it is the
competition between the energy associated with surface pairbreaking and the energy
cost of a domain wall separating two degenerate B-like phases.\cite{vor05}
Consider the two trajectories (labelled 1 and 2)  shown in Fig. \ref{fig:scattering}.
The left panel shows a homogenous B-phase,
while the right panel shows two degenerate B-like phases corresponding to amplitudes
$-\Delta_{\perp}$ left of a domain-wall and $+\Delta_{\perp}$ to the right.

For the trajectory 1 that reflects from the free surface we have
$p_z\rightarrow -p_z$. This sign change is the origin of pairbreaking by a
specular surface in \He-B; it leads the suppression of $\Delta_{\perp}$ and
the formation of surface Andreev bound states.\cite{vor03} The energy cost
is directly related to the spectrum of surface states. By contrast the
shallow trajectory which passes through the dashed plane without
intersecting a surface encounters a nearly uniform order parameter
resulting in little or no pairbreaking.

Surface pairbreaking can be suppressed locally by compensating the sign change
that results from surface reflection. In particular, for the domain wall configuration
shown in the right panel of Fig. \ref{fig:scattering} the sign change for the scattering trajectory
($p_z\rightarrow -p_z$) is compensated by the sign change associated with the degenerate states
on opposite sides of the domain wall ($\pm\Delta_z$). Thus, over a few coherence lengths near
the domain wall, surface reflection does not lead to strong pairbreaking, and correspondingly
the energy cost of surface scattering is reduced. However, it is not all ``savings''. There is
an energy cost for the domain wall. The shallow trajectory crossing the domain wall
now incurs a sign change. Pairbreaking occurs near the domain wall and a spectrum of
Andreev bound states forms on the interface.

For very thick films ($D\gg D_{c}(T)$)
the translationally invariant B-phase is favored because the surface pairbreaking
energy is small compared with the pairbreaking cost of a domain wall.
But, for sufficiently thin films the domain wall energy is less than the surface
pairbreaking energy and the broken symmetry phase is favored. The critical line where
one domain wall is favored over the uniform B-phase is $D_{c1}(T)$.
For $D<D_{c1}$ multiple domains are favored. Further reduction in the film thickness
favors more domain walls until they dissolve into the P-phase at the $D_{c2}(T)$,
or a first-order transition to the A-phase occurs.

\begin{figure}[h]
\centerline{\includegraphics[width=0.85\hsize]{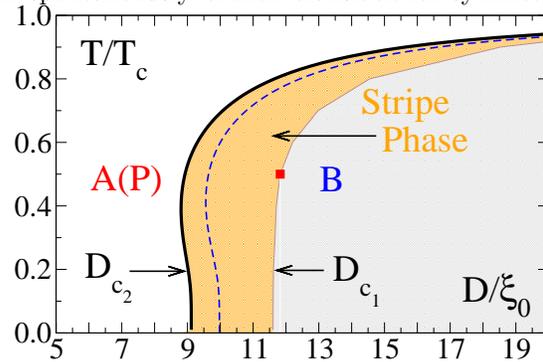}}
\caption{$T-D$-phase diagram for superfluid \He\ films in the weak-coupling
             limit ($p=0\,\text{bar}$). The phase with crystalline order separates
             translationally invariant $A(P)$ and $B$ phases.}
\label{fig:phase_diagram}
\end{figure}

Figure \ref{fig:phase_diagram} shows the $T$ vs. $D$ phase diagram for $^3$He films at
$p=0\,\text{bar}$. Calculations of the critical film thickness $D_{c1}(T)$ based on the
formalism described in Ref. \onlinecite{vor03} give a range of film thicknesses of order
$0.75\,\mu\text{m}\lesssim D \lesssim 1.0\,\mu\text{m}$ for $T\lesssim 0.75\,T_c$.

\par
In conclusion, our calculations predict that films of superfluid \He\
should exhibit an inhomogenous phase with spontaneously broken translational symmetry in
the plane of the film over a substantial range of temperatures and film thicknesses.
This phase has no analog in bulk \He, and should be identifiable by its anisotropic transport
properties. For example, the in-plane thermal conductivity should exhibit a reduced heat
conductivity normal to the direction of the stripes.
Signatures of the inhomogenous phase should also be observable as a broadening of
the NMR linewidth.



\end{document}